\documentclass[%
aip,
amsmath,amssymb,
reprint,%
]{revtex4-1}
\usepackage{graphicx}
\usepackage{color}
\usepackage{dcolumn}
\usepackage{bm}
\usepackage[utf8]{inputenc}
\usepackage[T1]{fontenc}
\usepackage{mathptmx}
\usepackage{etoolbox}
\makeatletter
\def\@email#1#2{%
 \endgroup
 \patchcmd{\titleblock@produce}
  {\frontmatter@RRAPformat}
  {\frontmatter@RRAPformat{\produce@RRAP{*#1\href{mailto:#2}{#2}}}\frontmatter@RRAPformat}
  {}{}
}%
\makeatother
\begin{document}
\title{Impact of Bending Stiffness on Ground-state Conformations for 
Semiflexible Polymers}
\author{Dilimulati Aierken}
\affiliation{Soft Matter Systems Research Group, Center for Simulational
Physics, Department of Physics and Astronomy, The University of Georgia,
Athens, GA 30602, USA}
%
\author{Michael Bachmann}
\homepage{https://www.smsyslab.org}
\affiliation{Soft Matter Systems Research Group, Center for Simulational
Physics, Department of Physics and Astronomy, The University of Georgia,
Athens, GA 30602, USA}
\email{bachmann@smsyslab.org}
\date{\today}
\begin{abstract}
Many variants of RNA, DNA, even proteins can be considered semiflexible
polymers, where bending stiffness, as a type of energetic penalty, competes
with attractive van der Waals forces in structure formation processes. Here,
we systematically investigate the effect of the bending stiffness on
ground-state conformations of a generic coarse-grained model for
semiflexible polymers. This model possesses multiple transition barriers.
Therefore, we employ advanced generalized-ensemble Monte Carlo methods to
search for the lowest-energy conformations. As the formation of distinct
versatile ground-state conformations including compact globules, rod-like
bundles and toroids strongly depends on the strength of the bending
restraint, we also performed a detailed analysis of contact and distance
maps.
\end{abstract}
\maketitle
\section{Introduction}
Biomolecules form distinct structures that allow them to perform specific
functions in the physiological environment. Understanding the effects of
different properties of these conformations is crucial in many fields, such as
disease studies~\cite{Sagis2004} and drug design~\cite{drug2020}. With the
recent development of computational resources and algorithms, computer
simulations have become one of the most powerful tools for studies of
macromolecular structures. However, atomistic or quantum level modeling is  
still
limited by the computational power needed to properly describe complex  
electron
distributions in the system, not to mention the thousands of ``force
field'' parameters to be tuned in semiclassical 
models~\cite{quantum2019,Atomic2011,Bachmann2014Book}. Moreover, such models 
are so specific that their results
usually lack generality. Thus, coarse-grained polymer models have been widely
used in recent years. Focusing on few main features, while other less relevant
degrees of freedom are considered averaged out, provides a more general view  
at
generic structural properties of polymers.

Semiflexible polymer models play an important role as they allow for studies  
of
various classes of
biopolymers~\cite{daniel2013,Janke2015,Chen2018,Majumder21,Shirts2022}, for
which the bending stiffness is known to be one of the key factors to be  
reckoned
with in structure formation processes. Bending restraints help DNA strands  
fold
in an organized way enabling efficient translation and transcription
processes~\cite{DNApacking}. RNA stiffness affects self-assembly of virus
particles~\cite{RNAstiff2018}. In addition, protein stiffness has been found  
to
be an important aspect in enzymatic catalysis processes, where proteins  
increase
stiffness to enhance efficiency~\cite{ProteinStiff2019}.

The well-known Kratky-Porod or worm-like chain (WLC) model~\cite{WLC} has
frequently been used in studies of basic structural and dynamic
properties of semiflexible polymers. However, lack of self-interactions in 
this
model prevents structural transitions. In this paper, we systematically study
the competition between attractive interactions, which usually are caused by
hydrophobic van der Waals effects in solvent, and the impact of the bending
stiffness for ground-state conformations of a coarse-grained model for
semiflexible polymers by means of advanced Monte Carlo (MC) simulations.

Our study helps identify the conditions which allow semiflexible 
polymers to form distinct geometric structures closely knitted to their 
biological function. For example, sufficient bending strength of the polymer 
chain is necessary for the formation of toroidal shapes. Such conformations 
are relevant for stable DNA-protein complexes~\cite{bustamante1,kulic1}. 
Also, DNA spooled into virus capsids tends to form toroidal structures, which 
support both optimal accommodation of DNA in a tight environment and the fast 
release due to the tension built up inside the capsid~\cite{linse1,cb1}.

The paper is organized as follows: Semiflexible polymer model and  
simulation methods are introduced in Sec.~\ref{sec:modmeth}.
Results of energetic and structural analyses of lowest-energy 
conformations are discussed in Sec.~\ref{sec:results}.
The summary in Sec.~\ref{sec:sum} concludes the paper.
\section{Model and Methods}
\label{sec:modmeth}
\begin{table*}
\caption{\label{tab:energies}Lowest-energy conformations and corresponding
energy values obtained from simulations for the selected values of the
bending stiffness ranging from $\kappa = 0$ (fully flexible) to $\kappa =
19$.}
\begin{tabular}{c}
\includegraphics[width = \textwidth]{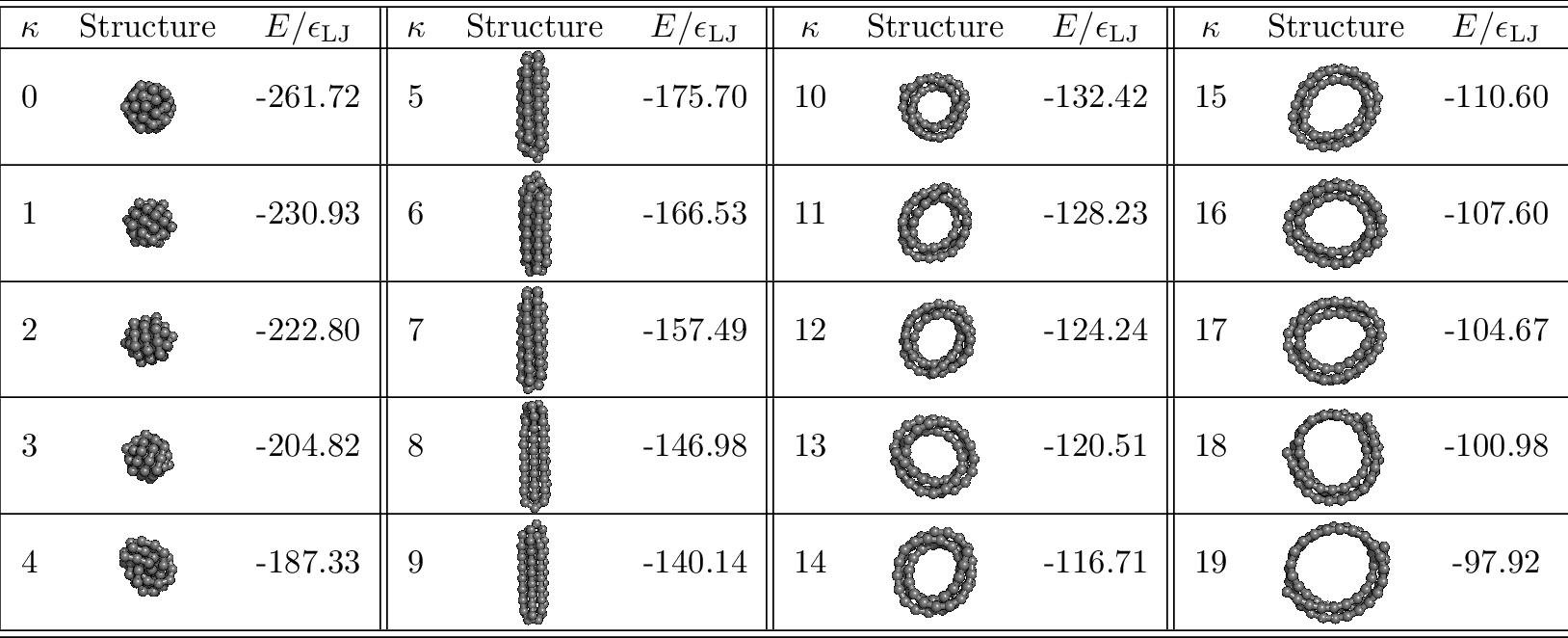}
\end{tabular}
\end{table*}
\subsection{Coarse-grained model for semiflexible polymers}
In a generic coarse-grained model for linear homopolymers, the monomers are
identical and connected by elastic bonds. Three energetic contributions are
considered in the model used in our study: bonded interactions, non-bonded
interactions and energetic penalty due to bending stiffness. The interaction
between non-bonded monomers, which depends on the monomer-monomer distance 
$r$,
\begin{equation}
V_{\mathrm{NB}}(r)=
\begin{cases}
V_{\mathrm{LJ}}(r)-V_{\mathrm{shift}}, & r<r_{c},          \\
0,                                     & \text{otherwise,}
\end{cases}
\end{equation}
is governed by the standard 12-6 Lennard-Jones (LJ) potential
\begin{equation}
V_{\mathrm{LJ}}(r)=4\epsilon_{\mathrm{LJ}}\left[\left(\dfrac{\sigma}{r}
\right)^{12}-\left(\dfrac{\sigma}{r}\right)^{6}\right].
\end{equation}
The energy scale is fixed by $\epsilon_{\mathrm{LJ}}$.
The potential minimum is located at $r_{0}=2^{1/6}\sigma $, where $\sigma$ is
the van der Waals radius. A cutoff at $r_{c}=2.5\sigma$ is applied to reduce
computational cost and the potential is shifted by a constant
$V_{\mathrm{shift}}\equiv V_{\mathrm{LJ}}(r_c)$ to avoid a discontinuity.

The bond elasticity between adjacent monomers is described by the combination 
of Lennard-Jones and finitely extensible nonlinear elastic (FENE)
potentials~\cite{Milchev2001, Kremer1990, Bird1987}, with the minimum located 
at $r_0$:
\begin{equation}
V_{\mathrm{B}}(r)=-\frac{1}{2}KR^{2}\mathrm{ln}\left[1-\left(\frac{r-r_{0}}{R}
\right)^{2}\right]+V_{\mathrm{LJ}}(r)-V_{\mathrm{shift}}.
\end{equation}
Here, the standard values $R=(3/7)r_{0}$ and
$K=(98/5)\epsilon_{\mathrm{LJ}}r_0^2$ are used~\cite{Qi2014}. Due to bond
rigidity, the fluctuations of the bond length $r$ are limited to the range
$[r_{0}-R, r_{0}+R]$.

To model the impact of chain rigidity, a bending potential is introduced. The
energetic penalty accounts for the deviation of the bond angle $\theta$ from 
the reference angle $\theta_0$ between neighboring bonds:
\begin{equation}
V_{\mathrm{bend}}(\theta)=\kappa\left[1-\cos(\theta-\theta_{0})\right],
\end{equation}
where $\kappa$ is the bending stiffness parameter. In this study we set 
$\theta _{0} = 0$.

Eventually, the total energy of a polymer chain with conformation 
$\boldsymbol{X}=\left(\boldsymbol{r}_1,...,\boldsymbol{r}_N\right)$ is given
by
\begin{equation}
E(\boldsymbol{X})=\sum_{i>j+1}V_{\mathrm{NB}}(r_{i,j})+\sum_{i}V_{\mathrm{B}}
(r_{i,i+1})+\sum_{l}V_{\mathrm{bend}}(\theta_{l}),
\end{equation}
where $r_{i,j}=|\boldsymbol{r}_i-\boldsymbol{r}_j|$ represents the distance
between monomers at positions $\boldsymbol{r}_i$ and $\boldsymbol{r}_j$.

The length scale $r_{0}$, the energy scale $\epsilon_{\mathrm{LJ}} $, as well  
as the Boltzmann constant $k_{\mathrm{B}}$ are set to unity in our 
simulations. The polymer chain consists of $N=55$ monomers~\cite{aierken20}.
\subsection{Stochastic Sampling Methods}
The model we have studied has a complex hyperphase diagram that 
exhibits a multitude of structural phases. Crossing the transition lines 
separating these phases in the search for ground-state conformations is a 
challenging task.
Advanced generalized-ensemble Monte Carlo (MC) techniques have been 
developed to cover the entire energy range of a system, including the 
lowest-energy states. In this study, we primarily used the replica-exchange 
Monte Carlo method
(parallel tempering)~\cite{Swendsen1986,Geyer1991,Hukushima1996_Japan,
Hukushima1996,Earl2005} and an extended two-dimensional version of 
it~\cite{Majumder21} with advanced MC update strategies.

In each parallel tempering simulation thread $k$, Metropolis Monte 
Carlo simulations are  
performed.
The Metropolis acceptance probability
that satisfies detailed balance is generally written as:
\begin{equation}
a(\boldsymbol{X} \to 
\boldsymbol{X}')=\mathrm{min}\left(\sigma(\boldsymbol{X},
\boldsymbol{X}')\omega(\boldsymbol{X},\boldsymbol{X}'), 1\right),
\end{equation}
where
$\omega(\boldsymbol{X},\boldsymbol{X}')=\exp\left(-(E(\boldsymbol{X}')-
E(\boldsymbol{X}))/k_{\mathrm{B}}T_k\right)$ is the ratio of microstate 
probabilities at temperature $T_k$, and
$\sigma(\boldsymbol{X},\boldsymbol{X'})=s\left(\boldsymbol{X'\to
X}\right)/s\left(\boldsymbol{X\to X'}\right)$ is the ratio of forward and
backward selection probabilities for specific updates. Replicas with the 
total energy $E_k$ and $E_{k+1}$ are exchanged between adjacent threads $k$ 
and $k+1$
with the standard exchange acceptance probability:
\begin{equation}
P=\mathrm{min}\left(\exp\left[\left(\beta_k - \beta_{k+1}\right)
\left( E_k-E_{k+1}\right)\right], 1 \right),
\end{equation}
where $\beta_k =(k_{\mathrm{B}}T_{k})^{-1}$ and 
$\beta_{k+1} =(k_{\mathrm{B}}T_{k+1})^{-1}$ are the corresponding inverse 
thermal energies. Displacement moves with adjusted 
box sizes for different temperatures were used to achieve about 50\% 
acceptance rate. A combination of bond-exchange moves~\cite{Schnabel2011}, 
crankshaft moves~\cite{Austin2018}, and rotational pivot updates helped to 
improve the sampling efficiency.

In order to expand the replica exchange simulation space, the total energy of
the system was decoupled,
\begin{equation}
E(\boldsymbol{X}) = E_0(\boldsymbol{X})+\kappa E_1(\boldsymbol{X}),
\end{equation}
where $E_0(\boldsymbol{X}) =
\sum_{i>j+1}V_{\mathrm{NB}}(r_{i,j})+\sum_{i}V_{\mathrm{B}}(r_{i,i+1})$ and
$E_1(\boldsymbol{X})=\sum_{l}\left[1-\cos(\theta_l-\theta_{0})\right]$. After
every $1500$ to $3000$ sweeps (a sweep consists of $N=55$ MC 
updates), replicas at neighboring threads  
$(T_k,\kappa_k)$ and $(T_{k+1},\kappa_{k+1})$ were proposed to be exchanged 
according to the probability~\cite{Majumder21}:
\begin{equation}
P_{\mathrm{ext}}=\mathrm{min}\left(\exp \left[ (\Delta \beta \Delta 
E_0)+\Delta(\beta \kappa)\Delta E_1 \right], 1 \right).
\end{equation}
Here $\Delta \beta = \beta_k-\beta_{k+1}$ and 
$\Delta(\beta\kappa)=\beta_{k}\kappa_{k}-\beta_{k+1}\kappa_{k+1}$.

In selected cases, optimization methods such as
Wang-Landau~\cite{WL,nonflatWL}, simulated annealing~\cite{SA} and 
Energy
Landscape Paving~\cite{ELP} were also employed to validate  
results obtained from the replica-exchange simulations.
\section{Energetic and Geometric Analysis of Putative Ground-state
Conformations} 
\label{sec:results}
\begin{figure}
\centering
\includegraphics[width = 0.5\textwidth]{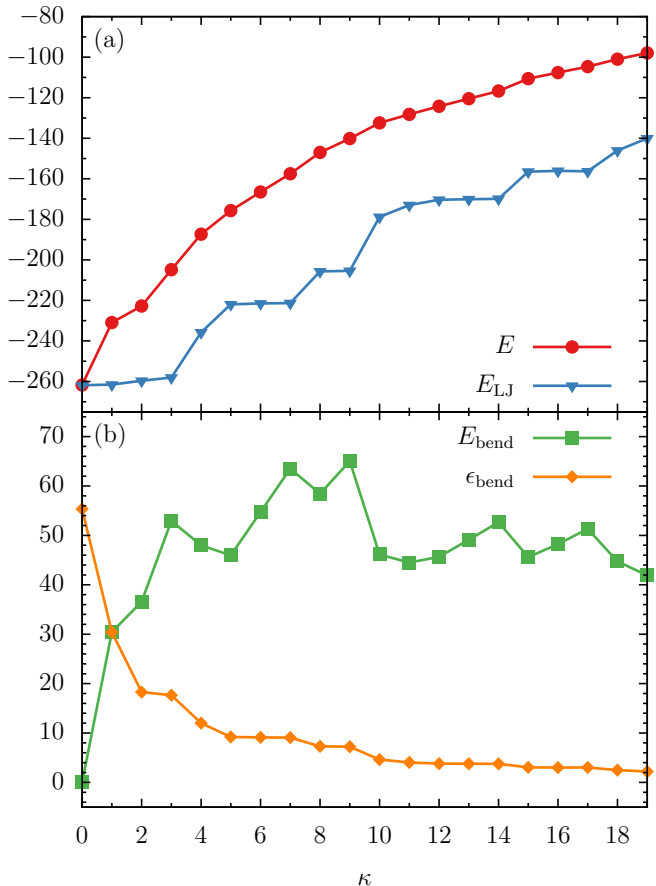}
\caption{(\textbf{a}) Total energy $E$ and Lennard-Jones contribution
$E_{\mathrm{LJ}}$ of ground-state conformations. \textbf{(b)} Total bending
energy $ E_{\mathrm{bend}}$ and renormalized bending contributions
$\epsilon_{\mathrm{bend}} = E_{\mathrm{bend}}/\kappa$ for the entire array
of $\kappa$ parameter values simulated.}
\label{fig:energy}
\end{figure}
In this section, we perform a detailed
analysis of the different energy contributions governing ground-state 
conformations of semiflexible polymers and discuss geometric properties based 
on the gyration
tensor. Eventually, we introduce monomer-distance and monomer-contact maps  
to investigate internal structural patterns.
\subsubsection{Energy Contributions}
Putative ground-state conformations and their energies obtained from
simulations for different choices of the bending stiffness $\kappa$ are listed
in Tab.~\ref{tab:energies}. By increasing the bending stiffness $\kappa$, the
semiflexible polymer folds into different classes of structures: compact
globules ($\kappa < 5$), rod-like bundles ($5 \leq \kappa \leq 9$), as well as
toroids ($\kappa > 9$). 

In order to better understand the crossover from one structure type to 
another, we first investigate the separate contributions  from LJ 
and bending potentials to the total ground-state energies. Since bond lengths 
are at almost optimal distances ($\approx r_0$), the bonded potential
$V_{\mathrm{FENE}}$ can be ignored in the following analysis. The main
competition is between
\begin{equation}
E_{\mathrm{LJ}} = \sum_{i>j} \left(V_{\mathrm{LJ}}(r_{i,j}) - 
V_{\mathrm{shift}}\right),
\end{equation}
including contributions from bonded monomers, and the bending energy
\begin{equation}
E_{\mathrm{bend}} = \sum_{l}V_{\mathrm{bend}}(\theta_{l}).
\end{equation}
We also introduce the renormalized contribution from the bending potential
\begin{equation}
\epsilon_{\mathrm{bend}} = E_{\mathrm{bend}}/\kappa
\end{equation}
for studying the relative impact of bending on these conformations.

The energies $E$,  $E_{\mathrm{LJ}}$, bending energy $E_{\mathrm{bend}}$, and
renormalized bending quantity $\epsilon_{\mathrm{bend}}$ are plotted for all
ground-state conformations in Fig. \ref{fig:energy}. Not surprisingly, the  
total energy $E$ increases as the bending stiffness $\kappa$ increases. 
Similarly, $E_{\mathrm{LJ}}$ also increases with increased bending stiffness 
$\kappa$, but rather step-wise. Combining these trends with the corresponding 
structures, it can be concluded that each major global change in ground-state 
conformations with increased bending stiffness leads to the reduced 
attraction between monomers (increase in $E_{\mathrm{LJ}}$). Whereas the 
bending energy $E_{\mathrm{bend}}$ does not exhibit a specific trend, the 
renormalized bending energy $\epsilon_{\mathrm{bend}}$ decreases step-wise as 
well for increased bending stiffness $\kappa$, as shown in 
Fig.~\ref{fig:energy}(b). It is more
interesting, though, to see there are clear alterations of $E_{\mathrm{LJ}}$  
and
$\epsilon_{\mathrm{bend}}$ within the same structure type (compact globules,
rod-like bundles, or toroids). 

In certain $\kappa$ intervals (e.g., $3<\kappa<5$ and $9<\kappa<10$), 
a rapid  
increase in $E_{\mathrm{LJ}}$ correlates with a decrease in  
$\epsilon_{\mathrm{bend}}$, which seems to be counter-intuitive. However, 
these are the regions, in which the structural type of the ground state 
changes significantly. This means a loss of energetically favorable contacts 
between monomers is not primarily caused by a higher bending penalty, but 
rather the global rearrangement of monomers.

For
$\kappa = 0, 1$ and $2$, the overall attraction $E_{\mathrm{LJ}}$ does not
change much, in contrast to $\epsilon_{\mathrm{bend}}$, suggesting that the
polymer chain is able to accommodate the bending penalty without affecting
energetically favorable monomer-monomer contacts.   

Even though the energetic analysis provides more information about the
competition between different energetic terms, conclusions about the  
structural
behavior are still qualitative. Therefore, a more detailed structural analysis
is performed in the following.
\subsubsection{Gyration Tensor Analysis}
\begin{figure}[t]
\centering
\includegraphics[width = 0.5\textwidth]{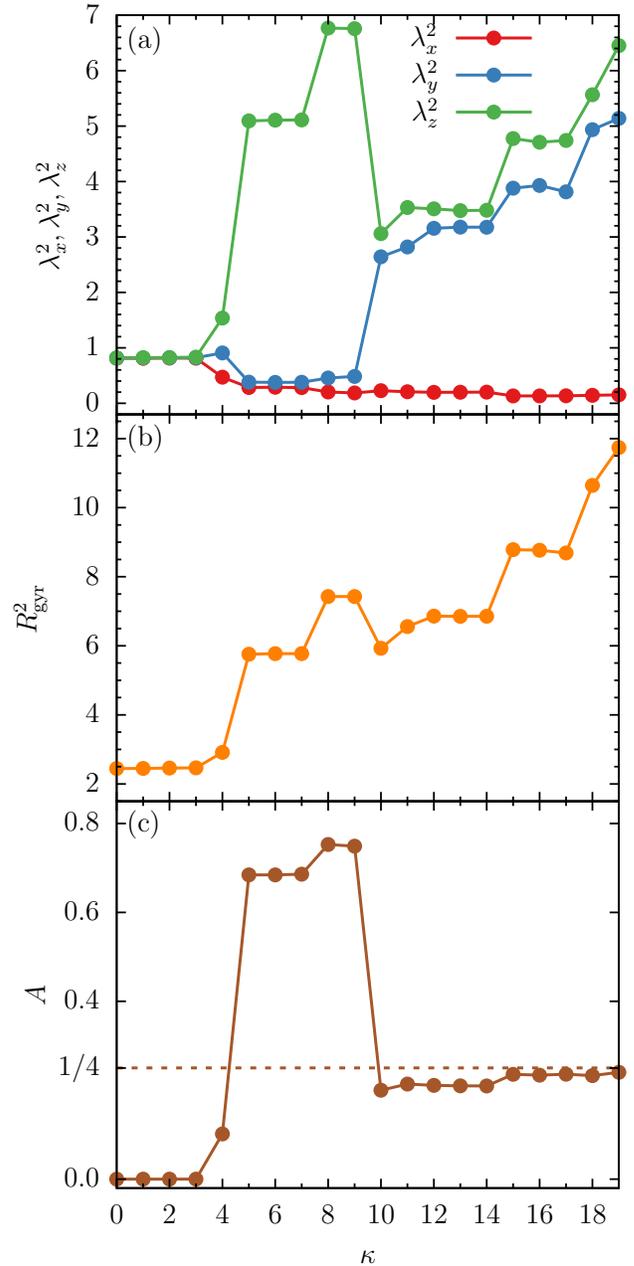}
\caption{(\textbf{a}) Square principal moments $\lambda_x^2, \lambda_y^2,
\lambda_z^2$ from the diagonalized gyration tensor $S$, (\textbf{b})
square radius of gyration $R^2_{\mathrm{gyr}}$, \textbf{(c)} and
relative shape anisotropy $A$ for ground-state conformations on a large
array of $\kappa$ values.}
\label{fig:tensor}
\end{figure}

In order to provide a quantitative description of the structural features, we
calculated the gyration tensor $S$ for the ground-state conformations with
components
\begin{equation}
S_{\alpha,\beta} = \frac{1}{N}\sum_{i=1}^{N} 
\left(r_\alpha^{(i)}-r_\alpha^{\mathrm{CM}}\right)\left(r_\beta^{(i)}-r_\beta^
{\mathrm{CM}}\right),
\end{equation}
where $ \alpha, \beta \in \left\{x,y,z\right\}$ and
$\boldsymbol{r}^{\mathrm{CM}} = \frac{1}{N}
\sum_{j=1}^{N} \boldsymbol{r}_j$ is the center of mass of the polymer.  After 
diagonalization, $S$ can be
written as
\begin{equation}
S_\mathrm{D} =
\begin{pmatrix}
\lambda^2_x & 0           & 0           \\
0           & \lambda^2_y & 0           \\
0           & 0           & \lambda^2_z
\end{pmatrix},
\end{equation}
where the eigenvalues are principal moments and ordered as $\lambda_x^2 \leq
\lambda_y^2 \leq \lambda_z^2$. These moments describe the effective extension  
of
the polymer chain in the principal axial directions. Thus, different invariant
shape parameters can be derived from combinations of these moments. Most
commonly used for polymers, the square radius of gyration  
$R^2_{\mathrm{gyr}}$
is obtained from the summation of the eigenvalues:
\begin{equation}
R^2_{\mathrm{gyr}} = \lambda^2_x+\lambda^2_y+\lambda^2_z.
\end{equation}
The radius of gyration describes the overall effective size of a polymer
conformation. In addition, another invariant shape parameter we employed is the
relative shape anisotropy $A$, which is defined as
\begin{equation}
A=\frac{3}{2}\frac{\lambda^4_x+\lambda^4_y+\lambda^4_z}%
{\left(\lambda^2_x+\lambda^2_y+\lambda^2_z\right)^2}-\frac{1}{2}.
\end{equation}
It is a normalized parameter, the value of which is limited to the interval 
$A\in [0,1]$, where $A=0$ is associated with spherically symmetric polymer 
chains
($\lambda_x = \lambda_y = \lambda_z$), and $A=1$ is the limit for the  
perfectly
linear straight chain ($\lambda_x = \lambda_y = 0, \lambda_z > 0$). Other than
these two limits, $A=1/4$ refers to perfectly planar conformations  
($\lambda_x =
0, 0 < \lambda_y = \lambda_z$). Square principal
components $\lambda_x^2, \lambda_y^2, \lambda_z^2$, square radius of gyration
$R^2_{\mathrm{gyr}}$, and the relative shape anisotropy $A$ of ground-state
conformations are plotted in Fig.~\ref{fig:tensor} as functions of $\kappa$.
\begin{figure*}
\centering
\includegraphics[width = 1.0\textwidth]{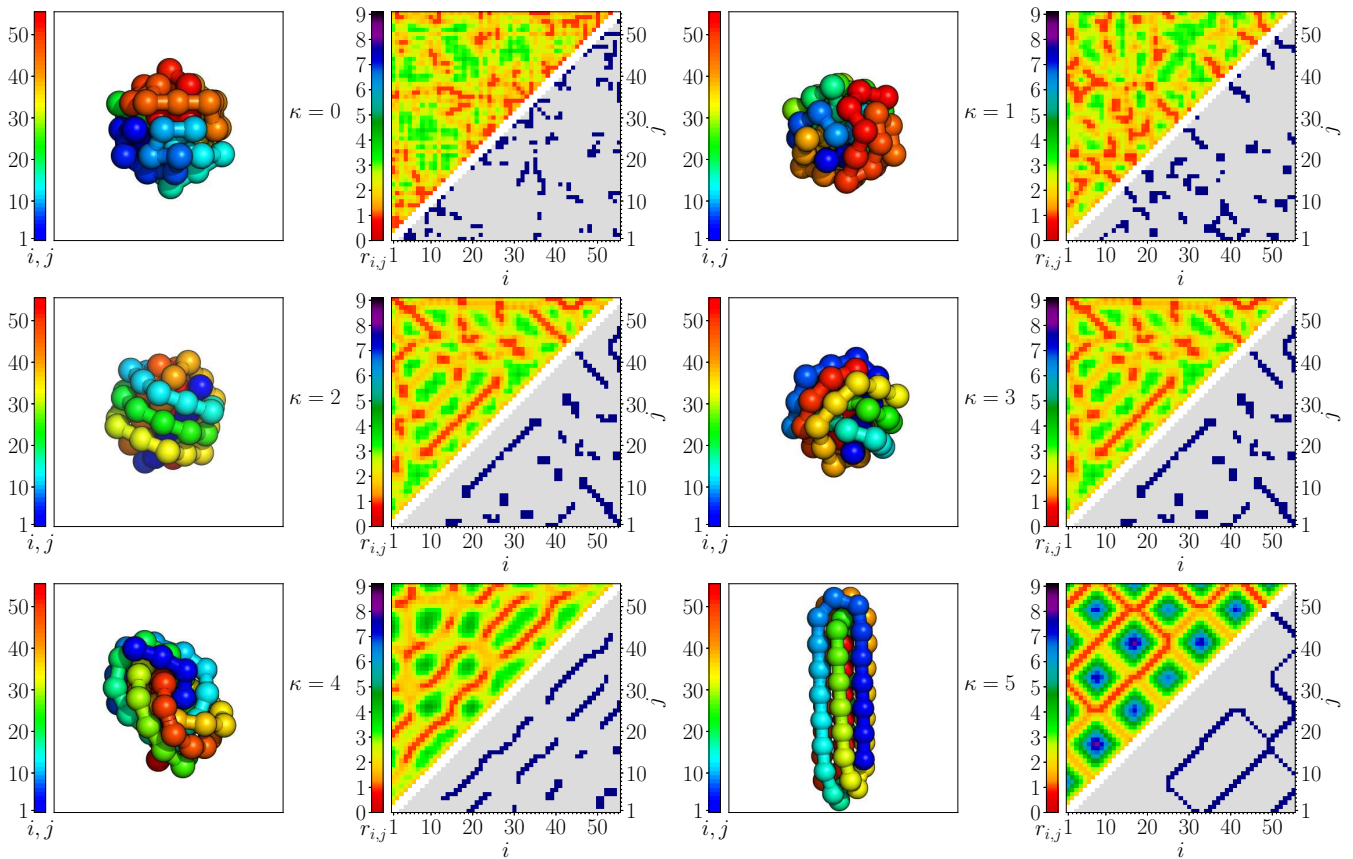}
\caption{Representations of ground-state conformations (left panel) and
their contact maps (right) for $\kappa \leq 5$. The upper triangle contains
the monomer distance map, where the distance $r_{i,j}$ of monomers $i$ and
$j$ is colored. The contact map is shown in the lower triangle. Monomers $i$
and $j$ are in contact if $r_{i,j} < 1.2$.}
\label{fig:map}
\end{figure*}

Starting with $\kappa = 0,1,2$ and $3$, the three principal moments of the
corresponding lowest-energy conformations are small and nearly equal. These  
are
the most compact conformations we found (see Tab.~\ref{tab:energies}). For 
these structures, $A< 10^{-3}$. Furthermore, for 
$\kappa<4$, lowest-energy conformations of semiflexible polymers 
possess an icosahedral-like arrangement of monomers, similar to that of 
the purely flexible chain ($\kappa=0$).

For $\kappa = 4$, the increased bending stiffness already forces 
conformations to stretch out noticeably. This is  reflected
by the imbalance of the principal moments. Consequently, $A$ is nonzero and 
the overall size of the conformations becomes larger as
$R^2_{\mathrm{gyr}}$ suggests.

If the bending stiffness is increased to $\kappa = 5, 6$ and $7$, rod-like
structures with $7$ bundles are formed to minimize the total energy. One
principal moment increases dramatically while the other two moments decrease.  
As
a result, $R^2_{\mathrm{gyr}}$ reaches a higher level, but remains almost
constant in this $\kappa$ range. The relative shape anisotropy climbs to $A 
\approx 0.69$, indicating
that the shape straightens out further.

The number of bundles reduces to six for $\kappa = 8$ and $9$, resulting in
longer rod-like structures. Both $R^2_{\mathrm{gyr}}$ and $A$ increase  
further,
the change of which is not visually obvious in Tab.~\ref{tab:energies},  
though.

With the bending energy even more dominant for $10 \leq \kappa \leq 14$,
the appearance of conformations changes significantly. Toroidal
structures with up to $4$ windings are energetically more favored than  
rod-like
bundles. Instead of forming a few sharp turns to
accommodate the bending penalty as in the bundled conformations, the polymer 
chain now takes on a rather dense
toroidal shape. Successive bending angles are comparatively small. In  
this
case, the two largest principal moments converge to an intermediate value. As  
a
consequence of the more compact structures, $R^2_{\mathrm{gyr}}$ decreases 
with
increased bending stiffness. The asphericity $A$ drops below the
characteristic limit $1/4$, reflecting the planar symmetry of the toroidal
structures.
\begin{figure*}
\centering
\includegraphics[width = 1.0\textwidth]{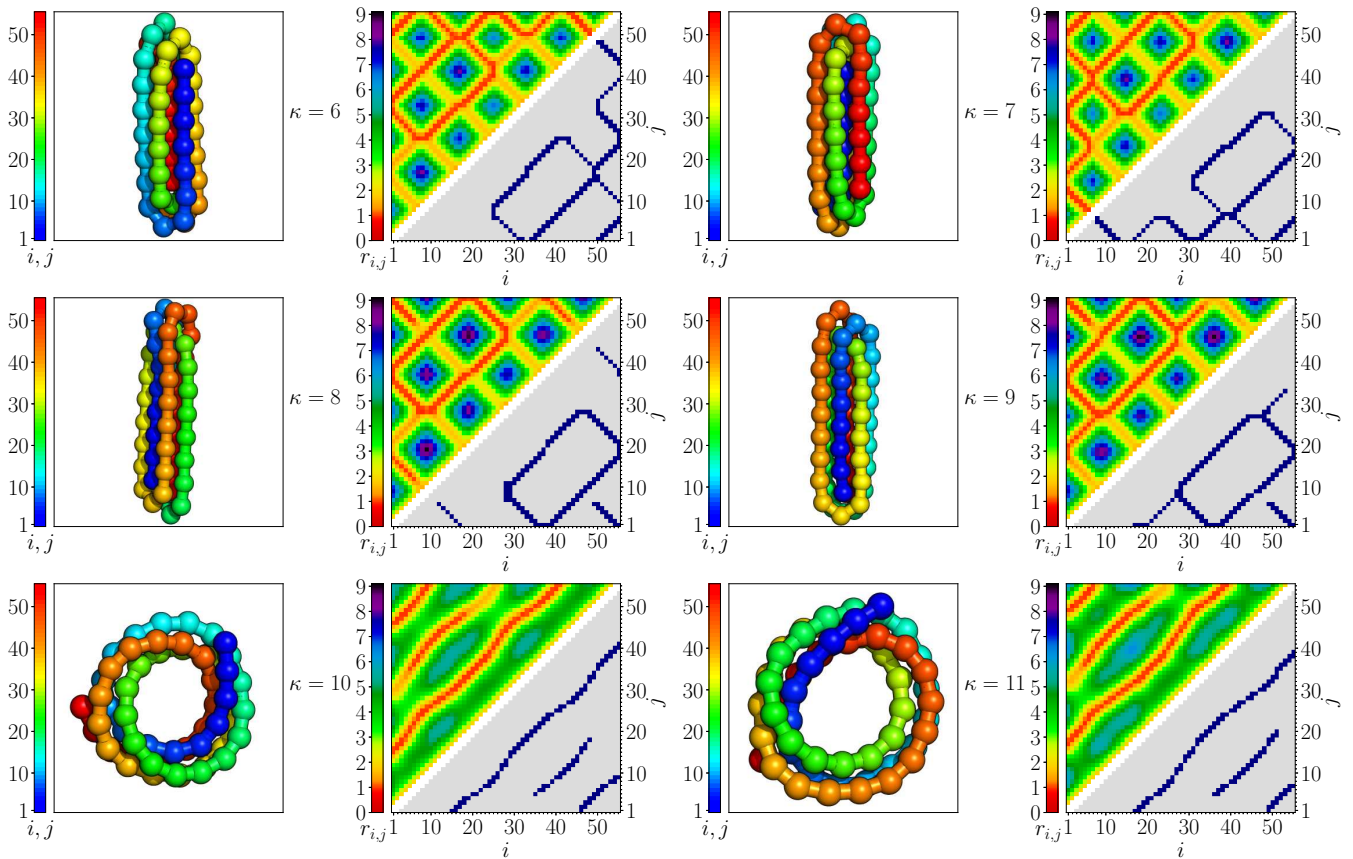}
\caption{Same as Fig.~\ref{fig:map}, but for $6 \leq \kappa \leq 11$.}
\label{fig:map2}
\end{figure*}
\begin{figure*}
\centering
\includegraphics[width = 1.0\textwidth]{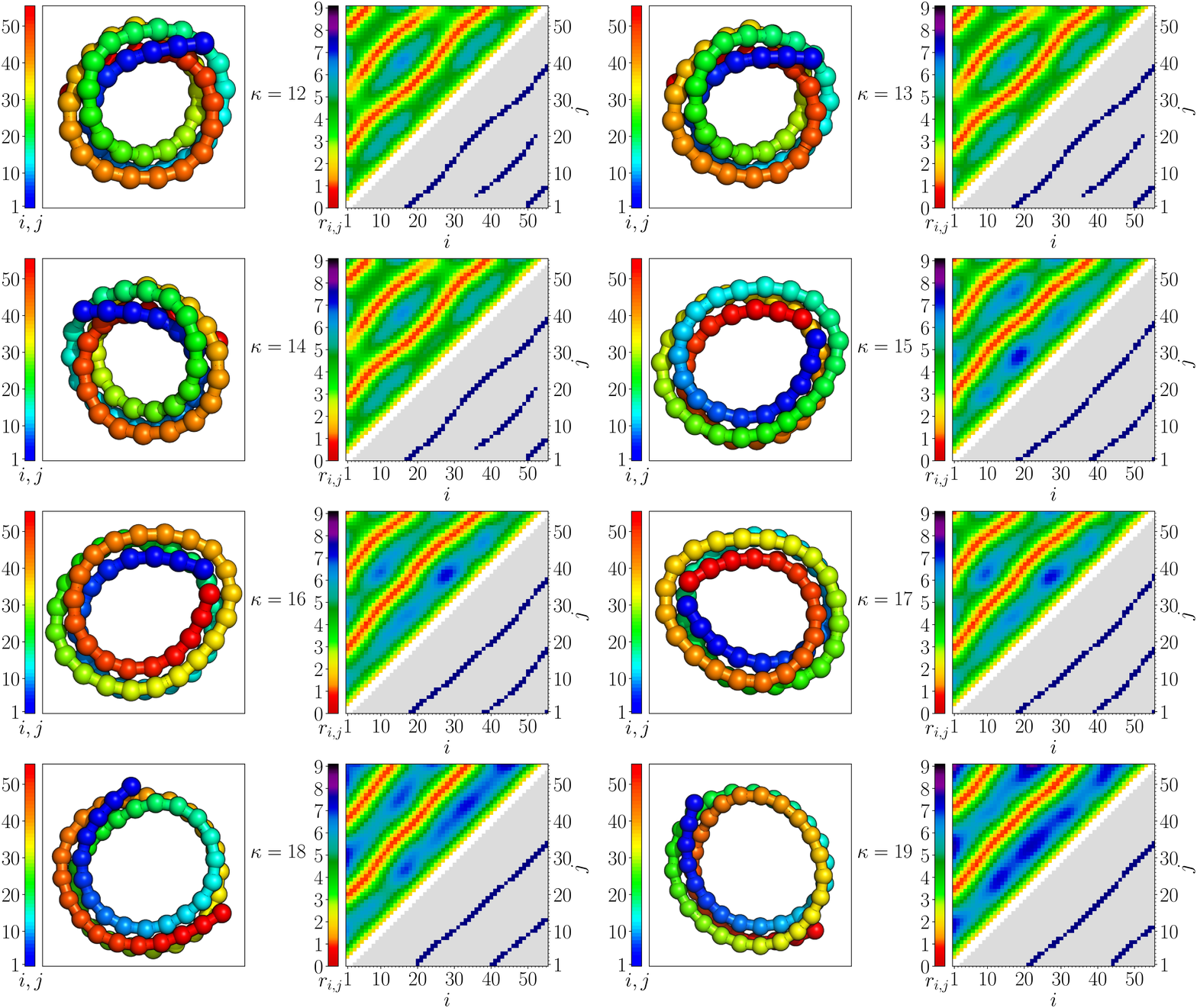}
\caption{Same as Fig.~\ref{fig:map}, but for $\kappa \geq 11$.}
\label{fig:map3}
\end{figure*}

It becomes more difficult for the polymer in the ground state to maintain the
same small bending angles for increased bending stiffness values $\kappa =
15,16$ and $17$. As a result, whereas the smaller bending angles still cause 
similar toroidal
structures as in the previously discussed case, the radius of the toroids 
increases and
fewer windings are present. Therefore, two main principal moments increase, 
as well as
$R^2_{\mathrm{gyr}}$. Meanwhile, the relative shape anisotropy $A$ approaches
$1/4$. Fewer windings reduce the overall thickness in the normal direction of
the toroidal conformations. As can be seen from the conformations in
Tab.~\ref{tab:energies}, these structures are stabilized by the attraction of
close end monomers.

However, for $\kappa > 17$, the attraction of two end monomers is not  
sufficient
to sustain the structure. Thus, expanding the toroid becomes an advantageous
option to offset strong bending penalties. The toroidal structure is stretched
out, which is clearly seen in Tab.~\ref{tab:energies} for $\kappa = 18$ and
$19$. The radius of the toroid keeps getting larger, so does
$R^2_{\mathrm{gyr}}$. We  
find
that $A$ keeps converging to the planar symmetry limit of $1/4$.

It is expected 
that increasing the bending stiffness further ultimately leads to a loop-like 
ground state and eventually to an extended chain, in which case no 
energetic contacts that could maintain the internal structural symmetries are 
present anymore. 
\subsubsection{Contact Map Analysis}
Even though the previous gyration tensor analysis yields a reasonable
quantitative description of the overall structural properties of the
ground-state conformations, it does not provide insight into internal
structures. Therefore, we now perform a more detailed analysis by means of
monomer distance maps and contact maps.
 
To find the relative monomer positions, we measured the monomer distance
$r_{i,j}$  between monomers $i$ and $j$ for all monomer pairs. Furthermore, we
consider nonbonded monomer pairs with distances $r_{i,j} < 1.2$ to be in
contact. The limit, which is close to the minimum distance $r_0$ of the 
Lennard-Jones
potential, allows to distinguish unique contact features of conformations  
while
avoiding counting nonnearest-neighbor contacts. In the figures, we colored the
monomers from one end to the other to visualize the chain orientation.

The combined results for $\kappa \leq 5$ are shown in Fig.~\ref{fig:map}. For
$\kappa = 0$ (flexible polymer), the structure is icosahedral, and the maps do
not exhibit particularly remarkable structural features. Without the energetic
penalty from bending, maximizing the number of nearest neighbors is the  
optimal
way to gain energetic benefit. For $\kappa = 1$, the introduced small bond  
angle
restraint already starts affecting the monomer positions. In the contact map,
short anti-diagonal streaks start appearing, which indicate the existence of
a U-turn like segment with two strands in 
contact.
Interestingly, we find similar conformations for $\kappa =2$ and $\kappa = 3$,
as confirmed by similar distance and contact maps. There are fewer, but longer
anti-diagonal strands, located in the interior of the compact structure. The
formation of new streaks parallel to the diagonal is associated with the  
helical
wrapping of monomers, which is visible in the colored representations. As for
$\kappa = 4$, the ground-state conformation is the compromise of two  
tendencies.
The bending stiffness neither is weak, as for $\kappa=3$ the semiflexible
polymer is still able to maintain a spherical compact structure with more  
turns,
nor is it particularly strong as for $\kappa = 5$, where the polymer forms a
rod-like bundle structure. Therefore, the lowest-energy conformations shown in
Fig.~\ref{fig:map} contain only helical turns trying to minimize the size, as
indicated by several diagonal streaks in the contact map. For $\kappa = 5$,  
the
polymer mediates the bending penalty by allowing only a few sharp turns  
between
the rods. For the $7$-bundle structure, the randomness completely disappears  
in
both distance and contact maps. The blue square areas in the distance map mark
the separation of monomer groups belonging to the two ends of a bundle. 
Furthermore, the diagonal streaks indicate the contact of two
parallel bundles while the turns of the chain form anti-diagonal streaks. It  
is
also worth mentioning that in this case the two end monomers are located on
opposite sides.

The results for $ 6 \leq \kappa \leq 11$ are shown in Fig.~\ref{fig:map2}.
Similar to $\kappa = 5$, the polymer still forms a $7$-bundle rod-like  
structure
for $\kappa =6$ and $\kappa = 7$. The anti-diagonal symmetry in maps for  
$\kappa
= 6 $ and $\kappa = 7 $ is only a consequence of opposite indexing of monomers.
For $\kappa = 8$ and $\kappa = 9$, the increased bending stiffness leads to a
decrease in the number of sharp turns from $7$ to $6$, where the two end
monomers are now located on the same side. The relative positions of monomers
are almost identical for $\kappa = 8$ and $\kappa = 9$ as seen in their  
distance
maps. However, the difference in contact maps is caused by the way the  
straight
rods following the sharp turns are aligned. For $\kappa = 8$, four monomers 
(the
orange turn in the colored presentation in Fig. \ref{fig:map2} for $\kappa =
8$) form the sharp turn. This allows the rods to align closer compared to the
$\kappa = 9$ case, where only $3$ monomers are located in the turn that holds
two parallel rods (blue shades). For $\kappa =10, 11$, the optimal way to pack
monomers is by toroidal wrapping. Thus, the contact maps exhibit only three  
diagonal
streaks.

Results for $\kappa \geq 11$ are shown in Fig.~\ref{fig:map3}. Contact maps  
for
$\kappa = 12, 13$ and $ 14$ still feature three diagonal streaks. However, for
$\kappa = 15, 16$, and $ 17$, the increased bending stiffness causes a larger
radius of the toroidal structure and the two end monomers are stabilized by
Lennard-Jones attraction. Thus, the number of parallel diagonals reduces to
two and the attraction of two end monomers is marked in the corners of the
maps. Finally, for polymers with even larger bending stiffness, i.e., 
$\kappa =18$ and $\kappa =19$, the contact between the two end monomers 
breaks and the whole
structure stretches out even more. As a result, the distance map for 
$\kappa =19$ contains extended sections of increased monomer distances. At the 
same time,
the contact map still shows two streaks slightly shifted to the right,
indicating a reduction in the number of contacts.
\section{Summary}
\label{sec:sum}
In this study, we have examined the effect of bending stiffness on  
ground-state
conformations of semiflexible polymers by using a coarse-grained model. In  
order
to obtain estimates of the ground-state energies, we employed an extended
version of parallel tempering Monte Carlo and verified our results by means  
of
global optimization algorithms. We find that the semiflexible polymer folds  
into
compact globules for relatively small bending stiffness, rod-like bundles for
intermediate bending strengths, as well as toroids for sufficiently large 
bending restraints.
Eventually, we performed energetic and structural analyses to study the impact
of the bending stiffness on the formation of ground-state structures.

We decomposed the energy contributions to gain more insight into the  
competition
between attractive van der Waals forces and the bending restraint. The total
energy of ground-state conformations increases smoothly with increased bending
stiffness, but not the attraction and bending potentials. Interestingly,
renormalizing the bending energy reveals that local bending effects of
ground-state conformations actually reduce for increased bending stiffness. 

The structural analysis by means of gyration tensor and invariant shape
parameters provided a general picture regarding the size and shape changes of
conformations under different bending restraints. In a further step, studying
distance maps and contact maps exposed details of internal structure ordering
and helped distinguish conformations, especially for small values of the  
bending
stiffness, where the gyration tensor analysis has been inconclusive. Contact  
map
analysis also caught slight differences, where different structure types are
almost degenerate.

In conclusion, the bending stiffness significantly influences the formation of
low-energy structures for semiflexible polymers. Varying the bending 
stiffness parameter in our model
results in shapes like compact globules, rod-like bundles, and toroids with
abundant internal arrangements. Semiflexible polymer structures remain
stable within a certain range of bending strengths, which makes them obvious
candidates for functional macromolecules. Monomer-monomer attraction provides
stability and bending stiffness adaptability to allow semiflexible polymers to
form distinct structures under diverse physiological conditions~\cite{ab23}.
\begin{acknowledgments}
This study was supported in part by resources and technical expertise from
the Georgia Advanced Computing Resource Center (GACRC).
\end{acknowledgments}
\section*{Data Availability Statement}
%
%
The data that support the findings of this study are available from the  
corresponding author upon reasonable request.

\end{document}